\documentclass[review]{elsarticle}
\usepackage{graphicx}
\usepackage{amsmath}
\usepackage{amssymb}
\usepackage{bm}
\usepackage{color}

\def\l{\langle}
\def\r{\rangle}
\begin{document}

\title{CUDA programs for GPU computing of Swendsen-Wang multi-cluster 
spin flip algorithm: \\
2D and 3D Ising, Potts, and XY models} 
\author[jaea,tmu]{Yukihiro Komura}
\ead{komura.yukihiro@jaea.go.jp}
\author[tmu]{Yutaka Okabe}
\ead{okabe@phys.se.tmu.ac.jp}
\address[jaea]{Nuclear Science and Engineering Directorate, 
Japan Atomic Energy Agency, 2-4 Shirakata-shirane, Tokai-mura,
Naka-gun, Ibaraki 319-1195, Japan}
\address[tmu]{Department of Physics, Tokyo Metropolitan University, Hachioji, Tokyo 192-0397, Japan}

\begin{abstract}
We present sample CUDA programs for the GPU computing of 
the Swendsen-Wang multi-cluster spin flip algorithm.  
We deal with the classical spin models; 
the Ising model, the $q$-state Potts model, and the classical 
XY model. As for the lattice, both the 2D (square) lattice and 
the 3D (simple cubic) lattice are treated. 
We already reported the idea of the GPU implementation for 2D models
[Comput. Phys. Commun. 183 (2012) 1155-1161].
We here explain the details of sample programs, 
and discuss the performance of the present GPU implementation 
for the 3D Ising and XY models.  We also show the calculated results of 
the moment ratio for these models, and discuss phase transitions. 
\end{abstract}


\maketitle

\noindent
{\bf Program summary}

\noindent
{\it Program title:} CUDA programs for GPU computing of Swendsen-Wang 
multi-cluster spin flip algorithm

\noindent
{\it Catalogue identifier:} AAAA

\noindent
{\it Program summary URL:} http://cpc.cs.qub.ac.uk/summaries/AAAA.html

\noindent
{\it Program obtainable from:} CPC Program Library, Queen's University, 
Belfast, N. Ireland

\noindent
{\it Licensing provisions:} Standard CPC licence, 
http://cpc.cs.qub.ac.uk/licence/licence.html

\noindent
{\it No. of lines in distributed program, including test data, etc.:} ****

\noindent
{\it No. of bytes in distributed program, including test data, etc.:} ****

\noindent
{\it Distribution format:} tgz

\noindent
{\it Programming language:} C, CUDA

\noindent
{\it Computer:} System with an NVIDIA CUDA enabled GPU 

\noindent
{\it External routines:} NVIDIA CUDA Toolkit 3.0 or newer

\noindent
{\it Nature of problem:}
Monte Carlo simulation of classical spin systems. Ising, 
$q$-state Potts model, and the classical XY model are treated 
for both two-dimensional and three-dimensional lattices.

\noindent
{\it Solution method:}
GPU-based Swendsen-Wang multi-cluster spin flip Monte Carlo method.
The CUDA implementation for the cluster-labeling is based on
the work by Hawick {\it et al.} 
[K.A. Hawick, A. Leist, and D. P. Playne, 
 Parallel Computing 36 (2010) 655-678],
and that by Kalentev {\it et al.} 
[O. Kalentev, A. Rai, S. Kemnitzb, and R. Schneider,
 J. Parallel Distrib. Comput. 71 (2011) 615-620].

\noindent
{\it Restrictions:}
The system size is limited depending on the memory of a GPU.

\noindent
{\it Running time:}
For the parameters used in the sample programs, it takes about a minute 
for each program.  Of course, it depends on the system size, 
the number of Monte Carlo steps, etc.

\section{Introduction}

Monte Carlo simulation is a standard method to study statistical physics 
of many body problems \cite{landau}.  The advance in computer hardware 
together with the development of software has enhanced 
the importance of Monte Carlo simulation all the time. 
The general purpose computing using the graphics processing unit (GPU) 
is a hot topic in computer science; 
using the common unified device architecture (CUDA) released by NVIDIA, 
it is now easy to implement algorithms on GPU 
using standard C or C++ language with CUDA specific extension. 
Preis {\it et al.} \cite{preis09,block12} proposed the efficient use 
of GPU for the Metropolis-type \cite{metro53} single spin flip 
Monte Carlo simulation of classical spin systems.  
Preis {\it et al.} \cite{preis09} reported the acceleration
of the computation 60 times for the two-dimensional (2D) Ising model, 
and 35 times for the three-dimensional (3D) Ising model 
compared to the CPU computation using a current CPU core. 
The parallel computing using a GPU 
is realized by interpenetrating sublattice decomposition, 
which is the same as vector computers in 1980s. 
Block {\it et al.} \cite{block10} further discussed the use of multiple GPUs 
for the Monte Carlo simulation as well as the multispin coding technique. 

The single spin flip Metropolis simulation often suffers from 
the problem of slow dynamics; that is, it takes long time 
for equilibration, for example, at temperatures near the critical 
temperature of the phase transition.  To conquer the problem 
of slow dynamics, the cluster spin flip algorithms of Monte Carlo 
simulation have been proposed.  The multi-cluster spin flip algorithm 
due to Swendsen and Wang \cite{sw87} and the single-cluster spin flip algorithm 
due to Wolff \cite {wolff89} are typical examples.  
The parallelization of cluster spin flip algorithm 
is not straightforward because the cluster labeling 
part of the cluster spin flip algorithm basically requires a sequential 
calculation, which is in contrast to the local calculation 
for the single spin flip algorithm.  The present authors \cite{komura12} 
proposed the GPU computing for the Swendsen-Wang (SW) multi-cluster spin flip 
algorithm, where the idea of Hawick {\it et al.} \cite{Hawick_labeling} 
was used in the cluster labeling part.  An improved version of 
the cluster labeling algorithm was proposed 
by Kalentev {\it et al.} \cite{Kalentev}. 
The present authors \cite{komura12} reported the acceleration of 
the GPU computing with 12.4 times for the SW algorithm of 
the 2D Ising model, in comparison 
with the CPU computing using a current CPU core.  
The present authors \cite{komura_multiGPU} extended the GPU computation 
for the SW multi-cluster spin flip algorithm to the multiple 
GPU computation, where two steps of parallelization are used, that is, 
the parallelization within each GPU and the inter-GPU parallelization. 
The multiple GPU computation of the multi-cluster spin flip algorithm 
was applied to the large-scale Monte Carlo simulation 
of the 2D classical XY model up to the system size $65536 \times 65536$ 
\cite{komura_XY}; the Kosterlitz-Thouless 
\cite{KT} transition temperature and the exponent to specify the multiplicative 
logarithmic correction were precisely determined.  
We note that the present authors also proposed the GPU computing 
for the Wolff single-cluster spin flip algorithm \cite{komura11}, 
although the efficiency of the acceleration was not so good 
as compared to the case of the multi-cluster spin flip algorithm. 

Here, we publicize sample CUDA programs for the GPU computing of 
the multi-cluster spin flip algorithm, and explain the details of 
the sample programs.  We deal with the following classical spin models; 
\begin{itemize}
\item 2D Ising model,
\item 2D Potts model,
\item 2D classical XY model,
\item 3D Ising model,
\item 3D Potts model,
\item 3D classical XY model.
\end{itemize}
As for the lattice, both the 2D (square) lattice and 
the 3D (simple cubic) lattice are treated. 

The Hamiltonian of the Ising model is given by
\begin{equation}
 \textrm{(Ising)} \quad
 \mathcal{H} = -J \sum_{\l i,j \r}s_{i}s_{j}, 
              \quad s_{i} = \pm 1.
\label{Ising}
\end{equation}
Here, $J$ is the coupling and $s_{i}$ is the Ising spin 
on the lattice site $i$. The summation is taken over 
the nearest neighbor pairs $\l i,j \r$.  
Periodic boundary conditions are employed. 
The $q$-state Potts model is an extension of the Ising 
model to the Potts spin with $q$ states, whose Hamiltonian is
\begin{equation}
 \textrm{(Potts)} \quad
 \mathcal{H} = -J \sum_{\l i,j \r}(\delta_{s_{i},s_{j}}-1), 
              \quad s_{i} = 1, 2, \cdots, q, 
\end{equation}
where $\delta_{a,b}$ is the Kronecker's delta. 
The 2-state Potts model is essentially the same as 
the Ising model except for the energy unit, but 
we separately give the program of the Ising model 
because it is easy for readers to start with the Ising model 
as a basic model. 
Finally, the Hamiltonian of the classical XY model, a continuous 
spin model, is given by
\begin{equation}
 \textrm{(classical XY)} \quad
 \mathcal{H} = -J \sum_{\l i,j \r} \bm{s}_i \cdot \bm{s}_j
   = - J \sum_{<i,j>} \cos(\theta_i - \theta_j),
\end{equation}
where $\bm{s}_i$ is a planar unit vector with two real components, 
$(\cos \theta_i, \sin \theta_i)$.  For actual implementation, 
the value of $\theta_i$ is discretized as $2\pi p_i/q$ with 
$p_i = 1, 2, \cdots, q$. This discretized model is referred to 
as the $q$-state clock model. 
When $q$ tends to infinity, the clock model becomes the classical XY model.
To make a cluster flip of vector spins, we use the idea of embedded 
cluster introduced by Wolff \cite{wolff89}.

The rest of the paper is organized as follows. 
In section 2, we briefly describe the SW multi-cluster 
spin flip algorithm and its GPU implementation for the Ising model 
as an example. 
In section 3, we report the performance of the present GPU implementation 
for the 3D Ising and XY models.  We show the calculated results of 
the moment ratio for these models, and discuss the phase transitions. 
We note that we already reported the performance of the 2D models 
in Ref. \cite{komura12}. 
We give the summary and discussion in section 4.

\section{GPU computation of Swendsen-Wang cluster algorithm}

In the SW multi-cluster spin flip algorithm \cite{sw87}, 
we flip clusters of spins at once. 
The SW algorithm for the Ising model, Eq. (\ref{Ising}), 
consists of three main steps: 
\begin{itemize}
\item[(1)] {\it Step of active bond generation}: Construct a bond lattice 
of active or non-active bonds with probability $p = 1-e^{2J/T}$, 
where $T$ is the temperature. 
\item[(2)] {\it Step of cluster labeling}: The active bonds partition 
the spins into clusters which are identified and labeled 
using a cluster-labeling algorithm.
\item[(3)] {\it Step of spin flip}: All spins in each cluster are set randomly to +1 or -1.
\end{itemize}

We should consider how to make parallel computation when we use GPUs.
Since the calculations of the step of active bond generation 
and the step of spin flip are done independently on each site, 
these steps are well suited for parallel computation on GPU. 
In the CPU computation of the step of cluster labeling,  
the Hoshen-Kopelman algorithm \cite{Hoshen_Kopelman}, 
which was first introduced in context of cluster percolation, 
is often used as an efficient algorithm. 
The Hoshen-Kopelman algorithm is a special version of 
the class of union-and-find algorithms \cite{cormen}. 
Since the assignment of label of cluster is done on each site 
piece by piece sequentially in the Hoshen-Kopelman algorithm, 
it is not straightforward to extend to parallel computing. 

Recently, Hawick {\it et al.} \cite{Hawick_labeling} 
investigated the cluster-labeling algorithm efficient 
for GPU calculation.  They proposed the labeling 
method of "Label Equivalence".  
The procedure of their algorithm was schematically explained 
in Fig. 1 of Ref. \cite{komura12}. 
In the CUDA program, the GPU "kernel" is invoked. 
The algorithm of Hawick {\it et al.} consists of 
three kernel functions, that is, 
scanning function, analysis function and labeling function, 
and two variables for labeling; one is a variable 
for saving the label, \verb+label+, and the other is 
a temporal variable for updated label, \verb+R+. 
The scanning function compares the label of each site with 
that of the nearest-neighbor sites when the bond between 
each site and the nearest-neighbor site is active. 
If the label of the nearest-neighbor site is smaller than the label 
of that site, the temporal variable with the label number, 
\verb+R[label[index]]+, is updated to the smallest one. 
For the update of the temporal variable on the scanning function, 
the atomic operation \verb+atomicMin()+ is used.  
Atomic operations provided by CUDA are performed 
without interference from any other threads. 
The analysis function resolves the equivalence chain of \verb+R+; 
the temporal variable \verb+R[index]+ 
is updated from the starting site to the new site. 
Each thread checks the temporal variable and the label on each site. 
When the label number, \verb+label+, is equal to the thread number, 
\verb+index+, each thread tracks back the temporal variable until 
the temporal variable, \verb+R+, remains unchanged. 
Since each thread executes this operation concurrently, 
the final value is reached quickly. 
The labeling function updates the label for saving 
by \verb+label[index]+ $\leftarrow$ \verb+R[label[index]]+. 
In the cluster-labeling algorithm of Hawick {\it et al.}, 
the loop including three functions is iterated up to the point 
when the information of the labeling needs 
no more process of scanning function. 

More recently, Kalentev {\it et al.} \cite{Kalentev} reported 
the refinement of the algorithm. 
The procedure of their algorithm is explained in Fig. 2 of Ref. \cite{komura12}.
They used only one variable for labeling instead of two 
because there is no need for a temporal reference; 
the implementation was improved in terms of memory consumption. 
It means that the number of kernel functions are reduced 
from three to two because the process of the labeling function 
is no more needed.  
They changed the execution condition on the analysis function, 
and eliminated the atomic operation. 
With the refinements due to Kalentev {\it et al.}, the improvement of 
computational speed and the reduction of the memory usage were realized. 

Now we explain our program \verb+AAAA+ in detail. 
The system size is represented by \verb+nx*nx+ for 2D models and 
\verb+nx*nx*nx+ for 3D models, where \verb+nx+ should be 
a multiple of 32.  As for 2D models, the maximum size 
of \verb+nx+ is 4096 or so, 
depending on memory size of GPU. 
If we use GTX580, the maximum linear size is 4096, whereas 
that is 8192 if we use GTX680, for example.  
As for 3D models, the maximum linear size is 256 for GTX580, 
whereas that is 512 for GTX680.
For the cluster-labeling algorithm, we can choose either 
the algorithm of Hawick {\it et al.} (\verb+algorithm = 0+), 
or the algorithm by Kalentev {\it et al.} (\verb+algorithm = 1+).
The number of Monte Carlo steps (MCSs) per spin for discard is 
denoted by \verb+mcs1+, whereas the number of MCSs 
per spin for measurement is denoted by \verb+mcs2+.

The kernel function 
\begin{eqnarray*}
  device\_function\_spinset;
\end{eqnarray*}
is a function for setting initial spin configuration.  
The kernel function for the step of active bond generation is
\begin{eqnarray*}
  device\_function\_init\_H;
\end{eqnarray*}
and
\begin{eqnarray*}
  device\_function\_init\_K;
\end{eqnarray*}
for the algorithm of Hawick {\it et al.} and 
for the algorithm of Kalentev {\it et al.}, respectively. 
Three kernel functions, 
\begin{eqnarray*}
  device\_function\_scanning\_H; \\ 
  device\_function\_analysis\_H; \\ 
  device\_function\_labeling\_H; 
\end{eqnarray*}
are used in the step of cluster labeling for the algorithm 
of Hawick {\it et al.}, whereas two kernel functions 
\begin{eqnarray*}
device\_function\_scanning\_K; \\ 
device\_function\_analysis\_K; 
\end{eqnarray*}
are used in the step of cluster labeling for the algorithm 
of Kalentev {\it et al.} 
The kernel functions 
\begin{eqnarray*}
device\_function\_spin\_select; \\
device\_function\_spin\_flip;
\end{eqnarray*}
are used in the step of spin flip. 
The kernel function 
\begin{eqnarray*}
device\_function\_sum; 
\end{eqnarray*}
is a function for measuring 
physical quantities.

For this sample program, 
we measure the magnetization 
\begin{equation}
 M = \sum_{i} s_{i}, 
\end{equation}
and the total energy, Eq. (\ref{Ising}).  In the calculation of total sum,
reduction of data is important to avoid data confliction. 
For the outputs, we print the 2nd moment of magnetization, 
the 4th moment of magnetization, the total energy per spin, 
and the specific heat per spin. 
We also print the computational time for spin-flip
and the computational time for spin-flip together with measurement 
of magnetization and energy.  The computation time is represented 
in units of milliseconds per a single MCS.
It is easy to extend our sample program to calculate other quantities, 
such as the correlation function, the helicity modulus, etc. 

We make comments on the technical part. 
The proper use of shared memories is important to save 
computational time. 
To improve the computational speed and save memory, we store 
the information on spin, bond and label in one word. 
Here, we use a linear congruential random-number generator which was 
proposed by Preis {\it et al.} \cite{preis09}, but any other 
random-number generators can be used.

A brief remark is made on the $q$-state Potts model. 
The program of the Potts model is essentially the same as that 
of the Ising model except that the number of states is $q$ 
instead of 2.  
The square of the magnetization of the $q$-state Potts model is 
calculated as 
\begin{equation}
   M^2 = \frac{q \ \sum_{k=1}^q n[k]^2-N^2}{q-1},
\end{equation}
where $n[k]$ is the number of spins with the state $k$, and 
$N$ is the total number of spins. 

Next, we discuss the GPU-based calculation of SW multi-cluster 
algorithm for the XY model, actually the $q$-state clock model. 
To make a cluster flip, we use the idea of 
embedded cluster introduced by Wolff \cite{wolff89}. 
We project vector spins to form Ising spin clusters. 
The essential part of the GPU implementation is the same 
as the case of the Ising model. 
The variable \verb+ispr+, which is randomly chosen, 
specifies the direction for mirror projection, and 
it is constant for all the sites, that is, all the threads.
We note that the proper use of shared memories is effective 
especially for the calculation of the inner product of vectors. 
The number of states $q$ should be less than or equal to 512 
because we use shared memory for the table of rule for 
energy, \verb+rule[ ]+, and the table 
for the magnetization, \verb+fmy[ ]+.

\section{Results}

We have tested the performance of our code on NVIDIA GeForce GTX580. 
Since the performance for 2D models was discussed in Ref. \cite{komura12}, 
we here show the data for 3D models. 

\begin{table*}[htbp]
\begin{center}
\begin{tabular}{llllll}
\hline
         &            & $L$=32    & $L$=64    & $L$=128   & $L$=256 \\
\hline
GTX580                    & update only   & 7.59 nsec & 3.52 nsec & 3.01 nsec & 3.10 nsec \\
\ \ Kalentev {\it et al.} & + measurement & 8.87 nsec & 3.90 nsec & 3.31 nsec & 3.33 nsec \\
\hline
\end{tabular}
\caption{\label{tb:GPU_time_Ising}
Average computational time 
per a spin flip for the 3D Ising model at $T = 4.5115$.  
The time for only a spin
update and that including the measurement of energy and magnetization are
given.}
\end{center}
\end{table*}

We first deal with the 3D Ising model. 
For the cluster-labeling algorithm, we show the data using 
the algorithm due to Kalentev {\it et al.} \cite{Kalentev}, 
because it is always faster than the calculation using 
that due to Hawick {\it et al.} \cite{Hawick_labeling}. 
We calculated the 3D Ising model near the critical temperature, 
$T = 4.5115$, \cite{ferrenberg,blote}. Throughout the present 
paper, we measure the temperature in units of $J$. 
The average computational times per a spin update 
near the critical temperature for the Ising model 
are tabulated in Table \ref{tb:GPU_time_Ising}. 
There, the time for only a spin update and 
that including the measurement of energy and magnetization are given.
We show the measured time per a spin flip in units of nano sec. 
The linear system sizes $L$ are $L$=32, 64, 128 and 256. 
We can see from Table \ref{tb:GPU_time_Ising} that the computational time 
of our GPU implementation of the SW algorithm is almost constant 
for $L \ge 128$; for smaller lattices the parallelization 
is not so effective.  
The performance for the 2D Ising model with the algorithm  
of Kalentev {\it et al.} was 2.51 nano sec per a spin flip 
with $L=4096$ on GTX580 \cite{komura12}.
It means that the computational time for a spin flip 
of the 3D model is compatible with that of the 2D model  
although the nearest neighbor sites are 6 instead of 4 
for the simple cubic lattice.

\begin{figure}
\begin{center}
\includegraphics[width=8.0cm]{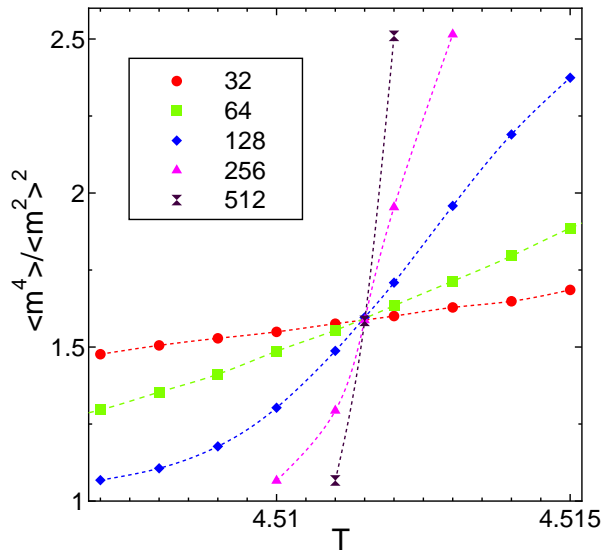}
\caption{\label{fig:fig1} 
Moment ratio of 3D Ising model; $L$=32, 64, 128, 256, and 512.}
\end{center}
\end{figure}

We next discuss the temperature dependence of physical quantities 
of the 3D Ising model.  It is known that this model undergoes 
the 2nd-order phase transition.  
The linear system sizes are $L$ = 32, 64, 128, 256, and 512. 
We discarded the first 10,000 MCSs, and 
the next 200,000 MCSs were used for measurement 
for $L \le 128$.  The MCSs for measurement 
were 100,000 and 50,000 for $L=256$ and for $L=512$, respectively. 
We used GeForce GTX680 for $L=512$. 
We plot the moment ratio 
\begin{equation}
   U(T) = \frac{\l M(T)^4 \r}{\l M(T)^2 \r^2}
        = \frac{\l m(T)^4 \r}{\l m(T)^2 \r^2}
\end{equation}
with $m=M/N$, which is essentially the Binder ratio \cite{Binder} 
except for the normalization, 
of the 3D Ising model in Figure \ref{fig:fig1}.  
We made four independent runs for each size; the average was 
taken over four runs, and the statistical errors were estimated, 
which are smaller than the size of the marks in Figure \ref{fig:fig1}. 
The crossing of the data with different sizes reproduces 
the known result of the critical temperature. 

\begin{figure}
\begin{center}
\includegraphics[width=8.0cm]{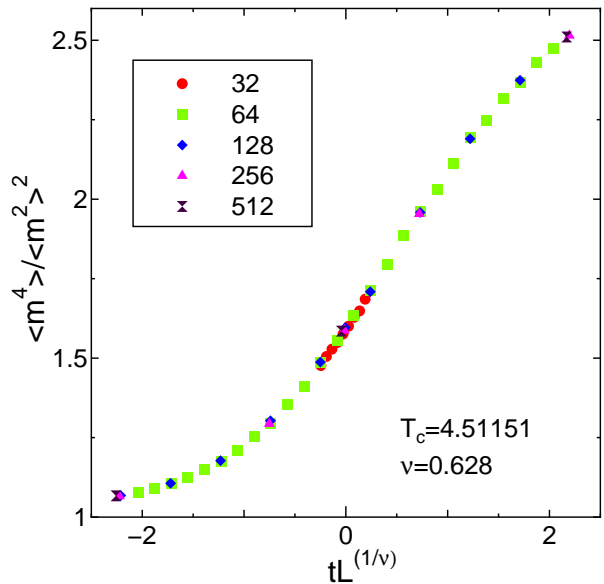}
\caption{\label{fig:fig2} 
FSS plot of moment ratio of 3D Ising model, 
where $t=(T-T_c)/T_c$ with $T_c$=4.51151 and $\nu$=0.628.}
\end{center}
\end{figure}

Let us consider the finite-size scaling (FSS) \cite{fisher70} 
of the moment ratio $U(T)$. 
We plot $U(T)=\l m^4 \r/\l m^2 \r^2$ as a function of $tL^{1/\nu}$; 
all the data with different sizes are collapsed on a single curve. 
Here, $t=(T-T_c)/T_c$, and $\nu$ is the critical exponent 
for the correlation length.  With the proper choice of $T_c$ and $\nu$, 
we can estimate $T_c$ and $\nu$. 
The FSS plot for the 3D Ising model is given in Figure \ref{fig:fig2}. 
The FSS is extremely good; a wrong value of the estimated 
$T_c$ yields a large deviation of FSS for large sizes. 
That is, we can get precise estimates with the data of large sizes. 
Here, the best choices of $T_c$ and $\nu$ are 4.51151(1) and 0.628(4), 
respectively. 
The previous estimates of $T_c$ are 4.511424(53) (=1/0.2216595(26)) 
(Ref. \cite{ferrenberg}) and 4.511524(20) (=1/0.2216546(10)) 
(Ref. \cite{blote}).  
We conclude that our estimate using large sizes up to 512 
is consistent with that of \cite{blote} within the statistical errors,
whereas it is outside of the error bars from that of \cite{ferrenberg}. 
The detailed analysis will be left to a future work, where 
much larger sizes will be treated with multiple GPUs.

\begin{table*}[htbp]
\begin{center}
\begin{tabular}{llllll}
\hline
         &            & $L$=32    & $L$=64    & $L$=128   & $L$=256 \\
\hline
GTX580                    & update only   & 8.78 nsec & 4.62 nsec & 4.15 nsec & 3.96 nsec \\
\ \ Kalentev {\it et al.} & + measurement & 10.33 nsec & 5.27 nsec & 4.57 nsec & 4.49 nsec \\
\hline
\end{tabular}
\caption{\label{tb:GPU_time_clock}
Average computational time per a spin flip 
for the 3D XY model at $T = 2.20175$.  
The time for only a spin update 
and that including the measurement of energy and magnetization are given.}
\end{center}
\end{table*}

The average computational times per a spin update 
near the critical temperature, $T = 2.20175$, for the classical XY model 
are tabulated in Table \ref{tb:GPU_time_clock}. 
The previous estimates of the critical temperature were given 
in \cite{Gottlob,Butera,Lan}. 
Actually, we dealt with the 512-state clock model. The discritization 
level is small enough except for very low temperatures. 
Again, the time for only a spin update and 
that including the measurement of energy and magnetization are given.
We show the measured time per a spin flip in units of nano sec. 
The linear system sizes $L$ are $L$=32, 64, 128 and 256. 
We can see from Table \ref{tb:GPU_time_clock} that the computational time 
of our GPU implementation of the SW algorithm is almost constant 
for $L \ge 128$.  
The computational time for the XY model is longer than the case 
of the Ising model because it needs more calculation 
for the projection to Ising spins.  But such an extra 
computational time is only a little, which shows 
that our GPU implementation of the SW multi-cluster algorithm 
is also effective for the vector spin model, the clock model.

\begin{figure}
\begin{center}
\includegraphics[width=8.0cm]{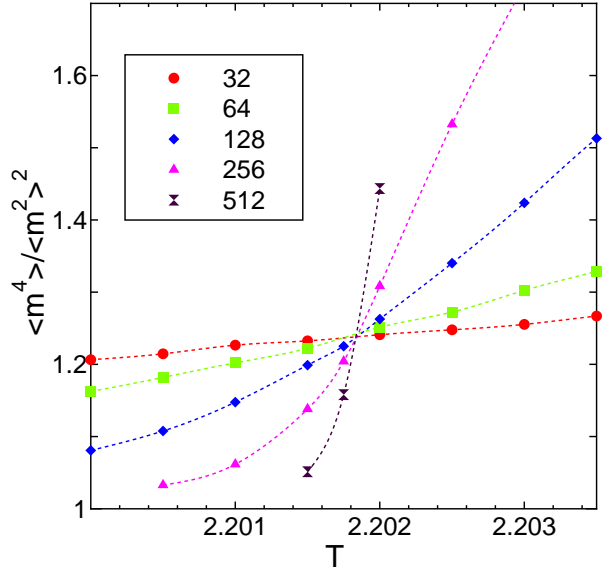}
\caption{\label{fig:fig3} 
Moment ratio of 3D XY model; $L$=32, 64, 128, 256, and 512.}
\end{center}
\end{figure}

We plot the moment ratio 
of the 3D XY model in Figure \ref{fig:fig3}.  
The linear system sizes are $L$ = 32, 64, 128, 256, and 512. 
The number of the MCSs for discard and measurement are 
the same as the case of the Ising model.
We made four independent runs for each size; the average was 
taken over four runs. 
The crossing of the data with different sizes again reproduces 
the known results of the critical temperatures.

\begin{figure}
\begin{center}
\includegraphics[width=8.0cm]{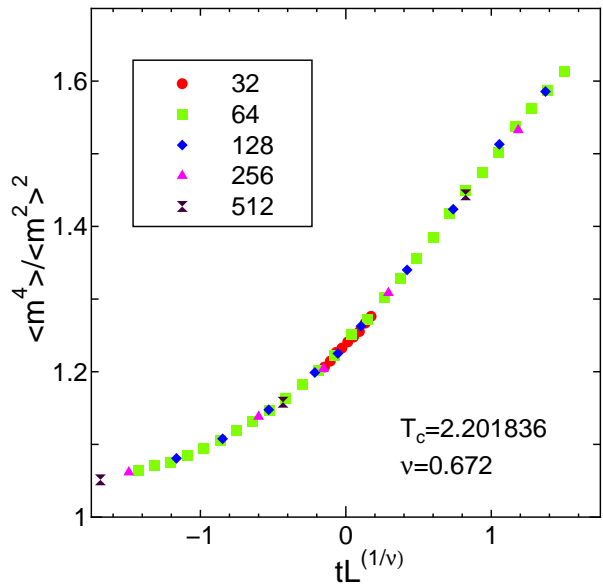}
\caption{\label{fig:fig4} 
FSS plot of moment ratio of 3D XY model, 
where $t=(T-T_c)/T_c$ with $T_c$=2.201836 and $\nu$=0.672.}
\end{center}
\end{figure}

We show the FSS plot of $U(T)=\l m^4 \r/\l m^2 \r^2$ as a function of 
$tL^{1/\nu}$ in Fig. \ref{fig:fig4}. 
Here, the best choices of $T_c$ and $\nu$ are 2.201836(6) 
and 0.672(4), respectively. 
The authors of recent publication \cite{Lan} gave two estimates of $T_c$ 
as 2.201852(1) and 2.2018312(6). 
Our estimate is between the two.  
The detailed analysis of the 3D classical XY model will also be left 
to a future work, where 
much larger sizes will be treated with multiple GPUs.

\section{Summary and discussion}

We have explained our program for the GPU-based SW multi-cluster 
spin flip algorithm for the classical spin models. 
Treated models are the Ising model, the $q$-state Potts model, 
and the classical XY model for 2D and 3D lattices. 
For the assignment of clusters, we used the algorithm 
by Hawick {\it et al.} \cite{Hawick_labeling} and 
that by Kalentev {\it et al.} \cite{Kalentev}. 
We have shown the performance of the 3D Ising model 
and the 3D classical XY model. 
The performance of GPU calculation is good, which is 
the same as the 2D case \cite{komura12}. 
Calculating the moment ratios of the 3D Ising and 
classical XY models, the critical temperature $T_c$
and the correlation-length exponent $\nu$ are 
discussed by using the FSS analysis. 
The system sizes we treated are up to $L=512$ for 
3D systems, but much larger sizes can be treated 
if we use multiple GPUs \cite{komura_multiGPU}. 
In near future, we make mutiple-GPU calculations 
for getting precise estimates for 3D models. 

In this paper, we have shown sample programs for the GPU-based 
SW multi-cluster spin flip algorithm of classical XY models. 
For the physical quantities we only calculated the magnetization 
and the energy.  The extension to the calculation of correlation function, 
helicity modulus, etc. is straightforward. 
We can extend the GPU calculation to other cluster algorithms 
including the quantum Monte Carlo simulations. 
We hope more researchers show interest in the GPU calculations 
of the cluster algorithms.

\section*{Acknowledgment}
This work was supported by a Grant-in-Aid for Scientific Research from
the Japan Society for the Promotion of Science.

\end{document}